\documentclass[doublespacing]{elsart} 

\usepackage{amsfonts,amsmath,amssymb, natbib,lineno}
\usepackage{epsf, subfigure, verbatim, epsfig, amsmath}
\newtheorem{result}{Result}
\newcommand{\ds}{\displaystyle}

\begin{document}

\begin{frontmatter}



\title{General Mechanisms for Inverted Biomass Pyramids in Ecosystems}


\author[Wang,HWang]{Hao Wang},
\author[Morrison]{Wendy Morrison},
\author[Singh]{Abhinav Singh},
\author[Wang]{Howard (Howie) Weiss}

\address[Wang]{School of Mathematics, Georgia Institute of
Technology, Atlanta, GA 30332, USA}
\address[Morrison]{School of Biology, Georgia Institute of
Technology, Atlanta, GA 30332, USA}
\address[Singh]{School of Physics, Georgia Institute
of Technology, Atlanta, GA 30332, USA}

\corauth[HWang]{Corresponding author: Telephone: 1-404-894-3949;
Fax: 1-404-894-4409; \newline {\em \hskip 0.25in Email address:}
wanghao@math.gatech.edu.}

\begin{abstract}
Although the existence of robust inverted biomass pyramids seem
paradoxical, they have been observed in planktonic communities, and
more recently, in pristine coral reefs. Understanding the underlying
mechanisms which produce inverted biomass pyramids provides new
ecological insights, and for coral reefs, may help mitigate or
restore damaged reefs. We present three classes of predator-prey
models which elucidate mechanisms that generate robust inverted
biomass pyramids. The first class of models exploits well-mixing of
predators and prey, the second class has a refuge (with explicit
size) for the prey to hide, and the third class incorporates the
immigration of prey. Our models indicate that inverted biomass
pyramids occur when the prey growth rate, prey carrying capacity,
biomass conversion efficiency, the predator life span, or the
immigration rate of prey fish is sufficiently large. In the second class,
we discuss three hypotheses on how refuge size can impact the amount of
prey available to predators. By explicitly incorporating
a refuge size, these can more realistically model predator-prey interactions
than refuge models with implicit refuge size.
\end{abstract}


\end{frontmatter}

\section{Introduction}

The biomass structure is a fundamental characteristic of ecosystems
\citep{odu71}. The shape of biomass pyramids encodes not only the
structure of communities, but also integrates functional
characteristics of communities, such as patterns of energy flow,
transfer efficiency, and turnover of different components of the
food web \citep{odu71,rei81,del99}.

A trophic pyramid is a graphical representation showing the energy
or biomass at each trophic level in a closed ecosystem. Energy
pyramids illustrate the production or turnover of biomass and the
energy flow through the food chain, while biomass pyramids
illustrate the biomass or abundance of organisms at each trophic
level. When energy is transferred to the next higher trophic level,
typically only 10\% is used to build new biomass \citep{pau95} and
the remainder is consumed by metabolic processes. Hence, in a closed
ecosystem, each trophic level of the energy pyramid is roughly 10\%
smaller than the level below it, and thus inverted {\it energy}
pyramids cannot exist.

A standard biomass pyramid is found in terrestrial ecosystems such
as grassland ecosystems or forest ecosystems, where a larger biomass
of producers support a smaller biomass of consumers \citep{das01}.
Although they appear to be rare, inverted biomass pyramids exist in
nature. They have been observed in planktonic ecosystems
\citep{del99}, where phytoplankton maintain a high production rate
and are consumed by longer lived zooplankton and fish. Recently,
inverted biomass pyramids have also been observed in pristine coral
reefs in the Southern Line Islands and Northwest Hawaiian Islands \citep{fri02,san08}, where the benthic
coral cover provides refuge for prey fish (Figure \ref{Kingman}).
At least one prominent researcher suspects that an apparent inverted
biomass pyramid exists on a reef off the North Carolina coast, and he speculates
this is due to significant immigration of prey fish into the reef
\citep{hay08}. In this manuscript, we introduce three classes of
predator-prey models to study how inverted biomass pyramids can
arise via these three distinct mechanisms.

\section{Well-Mixed Mechanism}

Most predator-prey models implicitly assume that predators and prey
are well mixed, and many incorporate a Holling-type predation
response \citep{hol59a,hol59b}. Although the ``well mixed''
assumption is usually far from being satisfied when prey are
animals, it appears to be a reasonable assumption for
phytoplankton-herbivore interactions in aquatic ecosystems, and we
first discuss the existence of inverted biomass pyramids in this
setting.

We begin by considering the standard Lotka-Volterra predator-prey
model with mass-action predation response \citep{lot25,vol26},
described by the system
\begin{eqnarray}\label{equ1}
\frac{dx}{dt} &=& ax - bxy,\\\label{equ2} \frac{dy}{dt} &=& cbxy -
dy,
\end{eqnarray}
where
\begin{align*}
x&: \text{prey biomass density}, &y: \text{predator biomass density},\\
a&: \text{prey growth rate}, &b: \text{per capita predation rate},\\
c&: \text{biomass conversion efficiency}, &d: \text{predator death rate}.\\
\end{align*}
The interior equilibrium point $\ds
(x^*,y^*)=\left(\frac{d}{cb},\frac{a}{b}\right)$ is neutrally stable
(a center), at which the predator:prey biomass ratio is
\begin{equation}\label{ratio1}
\ds \frac{y^*}{x^*}=\frac{ac}{d}.
\end{equation}
The ratio is greater than 1 if and only if $ac>d$. We obtain our
first result in biomass pyramid theory:

\begin{result}\label{theory1}
For the model (\ref{equ1})-(\ref{equ2}), if $ac>d$ (the prey growth
rate multiplied by the conversion efficiency is greater than the
predator death rate), the biomass pyramid is inverted; otherwise,
the biomass pyramid is standard.
\end{result}

Result \ref{theory1} provides a rigorous foundation for the belief
expressed by some biologists that inverted biomass pyramids result
from the high growth rate of prey and low death rate of predators
\citep{del99}. Result \ref{theory1} further suggests that the
biomass conversion efficiency can significantly influence the shape
of the biomass pyramid.

We now incorporate a general well-mixed predation response into the
predator-prey model, which is described by the system
\begin{eqnarray}\label{equ3}
\frac{dx}{dt} &=& ax - f(x)y,\\\label{equ4} \frac{dy}{dt} &=& c
f(x)y - dy,
\end{eqnarray}
where
\begin{align*}
f(x)&: \text{predation response function}.
\end{align*}
At the interior equilibrium point $(\hat{x},\hat{y})$, the ratio
$\hat{y}/\hat{x}=a/f(\hat{x})$, where $f(\hat{x})=d/c$. Thus the
predator:prey biomass ratio is
\begin{equation}\label{ratio2}
\ds \frac{\hat{y}}{\hat{x}}=\frac{ac}{d}.
\end{equation}
This interior equilibrium point is attracting when the system
(\ref{equ3})-(\ref{equ4}) is eventually bounded and has no stable limit cycles.
The system is eventually bounded if there is a bounded region where
all solutions eventually enter into and stay in. Result \ref{theory1} remains valid for this
extended model whenever the interior equilibrium point is stable. Actually, whenever the prey grow
exponentially, Result \ref{theory1} is robust to variations in refuge-dependent predation patterns.

We now incorporate logistic prey growth into the preceding
predator-prey model, which is described by the system
\begin{eqnarray}\label{equ5}
\frac{dx}{dt} &=& ax\left(1-\frac{x}{K}\right) -
f(x)y,\\\label{equ6} \frac{dy}{dt} &=& c f(x)y - dy,
\end{eqnarray}
where
\begin{align*}
K&: \text{prey carrying capacity},
\end{align*}
and the predation functional response $f(x)$ is a strictly
increasing function. Any reasonable predation function must satisfy
this monotone condition, which all three Holling-type functions do.
The monotonicity implies that the inverse function $f^{-1}$ exists
\citep{wik}, and thus the $x$-component of the interior equilibrium
point can be solved from $cf(x)=d$ as $\tilde{x}=f^{-1}(d/c)$. The
biomass ratio at the interior equilibrium point
$(\tilde{x},\tilde{y})$ can be written as
\begin{equation}\label{ratio3}
\ds
\frac{\tilde{y}}{\tilde{x}}=\frac{ac}{d}\left[1-\frac{f^{-1}(d/c)}{K}\right].
\end{equation}
This formula modifies the biomass ratio in model
(\ref{equ1})-(\ref{equ2}) and model (\ref{equ3})-(\ref{equ4}) by the
factor $\ds 1-\frac{f^{-1}(d/c)}{K}$. This interior equilibrium
point is attracting when the predator-extinction equilibrium $(K,0)$ is unstable and
there exist no stable limit cycles. For instance, if $\ds f(x)=\frac{bx}{\eta+x}$
is a Holling type II functional response, then the interior
equilibrium point is globally attracting whenever $\ds \frac{\eta d}{b-d}<K<\frac{\eta(b+d)}{b-d}$.
Under the stability condition, we obtain the new result:

\begin{result}\label{theory2}
For the model (\ref{equ5})-(\ref{equ6}), if $\ds
\frac{ac}{d}\left[1-\frac{f^{-1}(d/c)}{K}\right]>1$, the biomass
pyramid is inverted; otherwise, the biomass pyramid is standard.
\end{result}

The new condition for the inverted biomass pyramid depends
additionally on the prey carrying capacity $K$. We see that the
predator:prey biomass ratio is an increasing function of the prey
growth rate ($a$), the conversion efficiency ($c$), and the prey
carrying capacity ($K$), while the biomass ratio is a decreasing
function of the predator death rate ($d$). As a conclusion, we have
the following result:

\begin{result}\label{theory3}
The increase of prey growth rate, the conversion efficiency, the
prey carrying capacity, or the predator life span facilitates the
occurrence of inverted biomass pyramids.
\end{result}

Result \ref{theory3} is robust whenever the predation function is an
increasing function of prey density. We will see in the next section
that the same relations hold for refuge-dependent predation
functions.

\section{Refuge Mechanism}\label{section3}

Seeking refuge from predators is a general behavior of most animals
in {\it natural} ecosystems \citep{cow97,sih97} where the refuge
habitats can include burrows \citep{cla93}, trees \citep{dil89},
cliff faces \citep{ber91}, thick vegetation \citep{cas91}, or rock
talus \citep{hol91}. Some ecologists even believe that refuges
provide a general mechanism for interpreting ecological patterns
\citep{haw93}, specifically the extent of predator-prey interactions
\citep{huf58,leg03,ros06}. Aquatic ecologists have recently observed
inverted biomass pyramids in pristine coral reefs, where the benthic
coral cover provides the refuge for prey fish \citep{fri02,san08}.

In \citet{sin08}, we needed to introduce a refuge with explicit size.
Although the Holling type III functional response
offers the prey a refuge at low population density \citep{mur75},
the refuge is only implicit, and one can not specify the size of the
refuge. Some authors include an explicit refuge size into their
models by multiplying the prey density by $1-r$, where $0\leq r<1$
is a proxy of the refuge size
\citep{mcn86,sih87,hau94,kar05,hua06,kar06,ko06}. This procedure has
two fundamental drawbacks. The first is that for these modified predation
response functions, the switch point, where the predation rate
starts to quickly increase, critically depends on both the proxy refuge
size and the proxy half-saturation constant (independent of the refuge size).
The latter dependence is undesirable. For our model,
it is important that the switch point be a function of only the refuge size.
The second drawback is that, unlike the Holling-type
functional responses which are mechanistically derived from basic
biological principles, we have seen no derivation in the literature and
we are unable to mechanistically derive these functional forms from basic
biological principles to incorporate a refuge.

We now introduce a family of predator-prey models with explicit
refuge size, which we call {\bf refuge-modulated predator-prey (RPP)
models}. An important feature of this family is that the switch
points for the functional responses depend solely on the size
of the refuge. We group these models into three classes, RPP Types
I, II, and III, depending on the mechanistic dependence of prey
availability for predators on the refuge size. All previous refuge
models assume the mechanism behind our Type I class.

\subsection{Refuge-Modulated Predator-Prey Models}

In our recent work \citep{sin08}, we modeled the biomass of fish in
coral reefs. Small fish find refuge in coral reefs by hiding in
holes where large predators cannot enter \citep{hix93}.
This field observation motivated us to incorporate a refuge into the
standard predator-prey model, where the coral reef refuge size
influences the pattern of predation response. We introduced the
following family of models:
\begin{eqnarray}\label{equ7}
\frac{dx}{dt} &=& ax\left(1-\frac{x}{K}\right) -
f(x,r)y,\\\label{equ8} \frac{dy}{dt} &=& c f(x,r)y - dy,
\end{eqnarray}
where
\begin{align*}
r&: \text{refuge size},\\
f(x,r)&: \text{refuge-dependent predation response},\\
\end{align*}
and $f(x,r)$ is a strictly increasing function of prey biomass
density $x$. For each fixed $r$, the function $f_r(x)=f(r, x)$ is
strictly increasing in $x$, and thus its inverse $f_r^{-1}$ exists.
We solve for the $x$-component of the interior equilibrium point
from $cf(x,r)-d=0$ as $\bar{x}=f_r^{-1}(d/c)$. The biomass ratio at
the interior equilibrium point $(\bar{x},\bar{y})$:
\begin{equation}\label{ratio4}
\ds
\frac{\bar{y}}{\bar{x}}=\frac{ac}{d}\left[1-\frac{f_r^{-1}(d/c)}{K}\right].
\end{equation}
For each fixed refuge size $r$, the relationships between the
biomass ratio and other parameters are the same as in the well-mixed
predator-prey models. This provides the robustness of Result
\ref{theory3}. Additional hypotheses are needed to determine the
relationship between the biomass ratio and the refuge size. Although
the field observation in \citet{hix93} suggests that prey fish hide
in refuge places from predators, it is still unclear how the refuge
regulates the prey availability for predators. In the next
subsection, we propose three hypotheses all motivated from biological considerations.

\subsection{Hypotheses on Refuge Effects}

We model three biological hypotheses on how prey
availability for predators depends on the refuge size (Figure
\ref{Hypotheses}). We call these models RPP (Refuge-modulated
Predator-Prey) Type I, Type II, and RPP Type III. All predation
functions depend on the maximum predation rate $b$, the refuge size
$r$, the minimum predation rate regulator $\beta$, and the slope
regulator $\xi$.

\vspace{2ex}

\noindent \textbf{RPP Type I}: This model assumes that the prey availability
for predators decreases as the refgue size increases. Prey hide in the refuge, but trade-off protection (i.e. increased survival) for a decrease in growth or reproduction due to lower quality resources within the refuge \citep{per95,gon03,rea07}. Thus an increase in the size of the refuge protects more of the prey and results in less prey available to the predator.
Thus, the prey availability for predators is the prey density outside the
refuge and $f(x,r)$ is a decreasing function of refuge size $r$.
We choose the following representive function that can be fitted to empirical data:
\begin{equation} f(x,r)=\frac{b}{1+\beta e^{-\xi(x-r)}}.
\end{equation}
The variable $x$ is the total prey density (per unit area) and thus
the prey availability for predators is $x-r$.

The parameter $\beta$
captures the minimum predation rate as follows: when no prey and no
refuge are available, the predation rate is $\ds \frac{b}{1+\beta}$,
which is the minimum predation rate. We should choose $\beta$ sufficiently large
such that $\ds \frac{1}{1+\beta}<<1$, since it is reasonable
to have a small predation rate when no prey are available. The
parameter $\xi$ determines the slope of the predation curve when $x$ is close to $r$. The
prey density at the interior equilibrium point is $\ds
\bar{x}=f_r^{-1}(d/c)=r-\frac{1}{\xi}\ln\left[\frac{1}{\beta}\left(\frac{bc}{d}-1\right)\right]$.
The interior equilibrium point only exists when $bc>d$. To see this,
if $bc\leq d$ and  $\beta>0$, then $\ds \frac{dy}{dt}<(bc-d)y\leq
0$, and thus predators go extinct since all solutions tend to the
boundary equilibrium point $(K,0)$. Biologically, when the maximum
predation rate multiplied by the conversion efficiency is less than
the predator death rate, one would expect that predators cannot
persist. Under the conditions that $\beta$ is sufficiently large and
$bc>d$, the term $\ds
\frac{1}{\xi}\ln\left[\frac{1}{\beta}\left(\frac{bc}{d}-1\right)\right]$
is negative. Hence, $\ds
\bar{x}=f_r^{-1}(d/c)=r-\frac{1}{\xi}\ln\left[\frac{1}{\beta}\left(\frac{bc}{d}-1\right)\right]
=\boxed{r+\frac{1}{\xi}\ln\frac{\beta d}{bc-d}>0}$ for
sufficiently large $\beta$.

The evidence for the presence of trade-offs (survival vs. growth or reproduction) with the use of refuges is plentiful \citep{lim90,per95,rea07}. However, we are aware of only one experimental example that shows a decrease in growth rate of the predator in response to the use of a refuge by the prey \citep{per95}.

\vspace{2ex}

\noindent \textbf{RPP Type II}: This model assumes that the prey
availability for predators is independent of the refuge size (in the
sense of density, per unit area), i.e. $f(x,r)$ is a constant
function of $r$. Prey biomass within the refuge may increase, but the
amount available to the predators remains the same. We choose 
our representative RPP Type II predation function:
\begin{equation} f(x,r)=\frac{b}{1+\beta e^{-\xi x}}.
\end{equation}
The variable $x$ is the prey density, and the parameters $b$,
$\beta$, and $\xi$ have the same meanings as in RPP Type I. The
$x$-component of the interior equilibrium point is $\ds
\bar{x}=f_r^{-1}(d/c)=\boxed{\frac{1}{\xi}\ln\frac{\beta
d}{bc-d}>0}$  for $bc>d$ and sufficiently  large $\beta$.

We know of no biological examples of RPP Type II. However, we believe that RPP Types I and III are the extremes of a continuum, suggesting that condition can exist where the prey available to predators is not affected by the area within the refuge.

\vspace{2ex}

\noindent \textbf{RPP Type III}: This model assumes that the prey
availability for predators increases as the refuge size increases.
This may occur when resources such as food and mating sites are available within the refuge, allowing the prey to increase in numbers until some limiting resource forces a number of the prey to emigrate from the refuge in search for new habitat. The number of immigrants should be positively related to refuge size. Thus, $f(x,r)$ is an
increasing function of $r$. Our representative RPP Type III
predation function looks quite similar to our Type I predation
function, but the parameters require different interpretations:
\begin{equation}
f(x,r)=\frac{b}{1+\beta e^{-\xi(x+r)}}.
\end{equation}
The variable $x$ is the exterior (out of refuge) prey density (per
unit area), and  $x+r$ is the total prey density (per unit area). This
model assumes that the refuge stores a substantial amount of prey
and constantly provides food to predators, and thus the prey
availability is the total prey density (per unit area), i.e. $x+r$.

For $bc>d$ and $\beta$ sufficiently large such that $\ds
\frac{1}{\xi}\ln\frac{\beta d}{bc-d}>0$, the $x$-component of the
interior equilibrium point is $\ds \bar{x}=f_r^{-1}(d/c)=\boxed{-r+\frac{1}{\xi}\ln\frac{\beta d}{bc-d}>0}$ for
$r<\bar{r}$. The threshold refuge size $\ds
\bar{r}=\frac{1}{\xi}\ln\frac{\beta d}{bc-d}>0$ for $bc>d$ and
sufficiently large $\beta$. Because the refuge size in the model is
measured by density (per unit area), it is biologically reasonable to
assume a threshold maximum value for the refuge size.

The Elk Refuge in Yellowstone National Park is one example of a RPP Type III. The Elk Refuge provides protection (and food) to the elk during winter increasing survival to 97\% \citep{lub04}. The surviving elk migrate out of the refuge and provide a source of food for predators in Yellowstone National Park and surrounding areas \citep{smi08}. Our RPP Type III is also analogous to spillover and larval export hypotheses in marine protected areas (MPA) \citep{war01}. MPA’s are areas of the ocean that are protected from fishing (i.e. man is the predator). The fish within these MPA’s are hypothesized to increase the number of fish (prey) available outside the protected area through two mechanisms. The first, spillover, occurs when adult fish become crowded within the MPA and immigrate into the surrounding area. The second occurs when the fish within the MPA increase their reproductive output, increasing the number of recruits available to surrounding areas (larval export). While support for the spillover hypothesis is present (though limited spatially), it is much harder to prove the benefits of larval export \citep{war01}.

\vspace{2ex}

We now make a couple of general remarks about the RPP-type
functional responses. We always assume that $f(0,r)>0$ and small,
i.e. for each fixed $r$, the predation rate at zero prey density is
positive, but minimal. For Holling-type responses, $f(0,r)=0$. We
believe our choice is reasonable, since when the main prey species are no
longer available, predators may temporarily switch to alternative
lower quality food sources \citep{war98}. Thus, one must choose $\beta$
sufficiently large such that $\ds \frac{1}{1+\beta}<<1$. If we fit
this predation function to empirical data, $\xi$ may need to be
chosen large, depending on the size of $\beta$. When the prey
availability is  high, $f(x,r)$ is close to the maximum
predation rate $b$. Mathematically, the refuge size $r$ solely determines
the shift of the predation curve.

\subsection{Dependence of Biomass Ratio on the Refuge Size}

In this subsection, we use (\ref{ratio4}) to analyze the effects of
the refuge size on the predater:prey biomass ratio. It is evident that the
biomass ratio in (\ref{ratio4}) is a decreasing function of
$f_r^{-1}(d/c)$.

\vspace{2ex}

For RPP Type I, the term $\ds
f_r^{-1}(d/c)=r+\frac{1}{\xi}\ln\frac{\beta d}{bc-d}$ is an
increasing function of the refuge size $r$. Thus, the predator:prey
biomass ratio at the interior equilibrium point is a decreasing
function of the refuge size $r$.

\vspace{2ex}

For RPP Type II, the predator:prey biomass ratio is independent of
the refuge size.

\vspace{2ex}

For RPP Type III, the term $\ds
f_r^{-1}(d/c)=-r+\frac{1}{\xi}\ln\frac{\beta d}{bc-d}$, is
decreasing as the refuge size $r$ increases. Thus, the predator:prey
biomass ratio at the interior equilibrium point is an increasing
function of the refuge size $r$.

The following results immediately follow from these observations:

\begin{result}\label{theory4}
For the model (\ref{equ7})-(\ref{equ8}), if $\ds
\frac{ac}{d}\left[1-\frac{f_r^{-1}(d/c)}{K}\right]>1$, the biomass
pyramid is inverted; otherwise, the biomass pyramid is standard.
\end{result}

\begin{result}\label{theory5}
For RPP Type I, the decrease of the refuge size facilitates the
occurrence of inverted biomass pyramids. For RPP Type II, the refuge
size has no effects on biomass pyramids. For RPP Type III, the
increase of the refuge size facilitates the occurrence of inverted
biomass pyramids.
\end{result}

As an illustrative example, data from Kingman and Palmyra
\citep{san08} suggests that the predator-prey biomass ratio is an
increasing function of the refuge size (equivalent to the benthic
coral cover), and thus the appropriate predation response function
is RPP Type III. RPP Type III may be biologically appropriate if increases
in refuge size either increase recruitment or increase the survival
of recruits \citep{shu84,doh85}. After the surviving recruits grow into
juveniles or adults, they leave the refuge and provide an increase in
the food available to the predators.

\section{Immigration Mechanism}

Reef ecologists observed significant immigration of prey fish in a North
Carolina reef \citep{hay08}. We consider two types of
immigration: (i) immigrating prey fish stay in the coral reef and
adapt to survive in the new habitat; (ii) immigrating prey fish
leave the coral reef if they are not eaten by hungry predators, i.e.
they provide additional food to predators but do not add to the
local prey population. In this section, we incorporate both types of
immigration into the Lotka-Volterra predator-prey model:

\medskip

\noindent(i)
\begin{eqnarray}\label{equ9} \frac{dx}{dt} &=& ax - bxy +
\iota,\\\label{equ10} \frac{dy}{dt} &=& cbxy - dy;
\end{eqnarray}
(ii)
\begin{eqnarray}\label{equ11}
\frac{dx}{dt} &=& ax - bxy,\\\label{equ12} \frac{dy}{dt} &=&
cb(x+\iota)y - dy;
\end{eqnarray}
where $\iota$ is the constant immigration rate. For case (i), the
predator:prey biomass ratio at the interior equilibrium point
$(\tilde{x},\tilde{y})$ is
\begin{equation}\label{ratio5}
\frac{\tilde{y}}{\tilde{x}}=\frac{ac}{d}+\iota\frac{c^2 b}{d^2}.
\end{equation}
For case (ii), the predator:prey biomass ratio at the interior
equilibrium point $(\hat{x},\hat{y})$ is
\begin{equation}\label{ratio6}
\frac{\hat{y}}{\hat{x}}=\frac{ac}{d-\iota cb}.
\end{equation}
In both immigration cases, the biomass ratios are increasing
functions of the immigration rate $\iota$. This remains true when we
incorporate these two immigration effects into Holling type or RPP
type models. As a conclusion, we obtain the following robust result:

\begin{result}\label{theory6}
The immigration of prey  facilitates the occurrence of inverted
biomass pyramids.
\end{result}

\section{Discussion}

We develop a theory of biomass pyramids.
Our major contributions can be summarized as follows. First, when
prey grow exponentially, the biomass pyramid is inverted if and only
if the prey growth rate multiplied by the conversion efficiency is
greater than the predator death rate. Second, the increase of prey
growth rate, the conversion efficiency, the prey carrying capacity,
or the predator life span robustly facilitates the development of
inverted biomass pyramids. Third, based on plausible biological
hypotheses, we introduce a new series of predator-prey models
(called RPP type models) which explicitly and naturally incorporates a prey
refuge. Fourth, depending on the nature of an ecosystem, the
occurrence of inverted biomass pyramids can be positively or
negatively related to, or independent of, the refuge size. Fifth,
the immigration of prey facilitates the occurrence of inverted
biomass pyramids.

We propose three new refuge-dependent predation functions with explicit refuge size,
which capture the three essential biological hypotheses on the refuge (Figure
\ref{Hypotheses}).
The three can be combined into one function
\begin{equation}
\ds f(x,r)=\frac{b}{1+\beta e^{-\xi[x-(2-i)r]}},
\end{equation}
where $i$ is the index of RPP type, that is, $i=1$ for RPP Type I,
$i=2$ for RPP Type II, and $i=3$ for RPP Type III.

Some, but not all, of the prey that hide in the refuge are available to predators. Thus, there should be a discount
rate for the refuge size in the predation function of either RPP
Type I (assume no prey in the refuge are available) or RPP Type III (assume all prey in the refuge
are available). We incorporate this discount rate into the general
refuge-dependent predation function:
\begin{equation}
\ds f(x,r)=\frac{b}{1+\beta e^{-\xi(x+\eta r)}},
\end{equation}
where $-1\leq \eta\leq 1$. This model is close to RPP Type I if
$-1\leq \eta<0$, close to RPP Type II if $\eta=0$, and close to RPP Type III
if $0<\eta\leq 1$. We call $\eta$ as the refuge-effect parameter.

What characteristics of the prey might lead to RPP Type I versus RPP Type III? As stated above, RPP Type I will occur when the use of the refuge results in strong trade-offs between survival and reproduction or growth. Most previous theoretical models assume that the hypothesis for RPP Type I is the case; however, we hypothesize that RPP Type III will occur when the prey have the ability to reproduce within the refuge and/or when the refuge increases prey survival through a population bottleneck.

Prey animals seek refuges to hide from predators and thus
it is sometimes necessary to explicitly incorporate the refuge mechanism into
the predation function of predator-prey models. The family
of RPP-type models explicitly incorporating the refuge size can more accurately
describe realistic predator-prey interactions in ecosystems. We believe that RPP-type models provide the next
generation of models for predator-prey interactions. In the coming Winter, we plan to test these models via microcosm
experiments.

\section*{Acknowledgement}

\noindent We would like to thank Mark Hay for insightful comments
and helpful discussions, Alan Friedlander and Bruce Smith for their useful
feedback and references to our questions. We also would like to thank Lin Jiang for
his suggestions and allowing us to perform refuge experiments in his
lab in the near future.

\newpage

\begin{figure}
\centerline{\hbox{\epsfig{figure=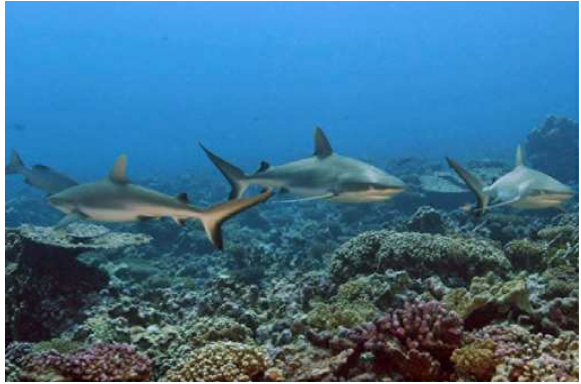,height=2in,width=3in,angle=0}}}
\caption{\label{Kingman} The pristine coral reef at Kingman.}
\end{figure}

Figure \ref{Kingman}. This figure is reproduced from \citet{san08}.
At Kingman coral reef, it was recently
discovered that apex predators constitute 85\% of the total fish
biomass. The biomass pyramid is clearly inverted in this pristine
coral reef. This is in sharp contrast to most reefs where the prey
biomass substantially dominates the total fish biomass.

\begin{figure}
\centerline{\hbox{\epsfig{figure=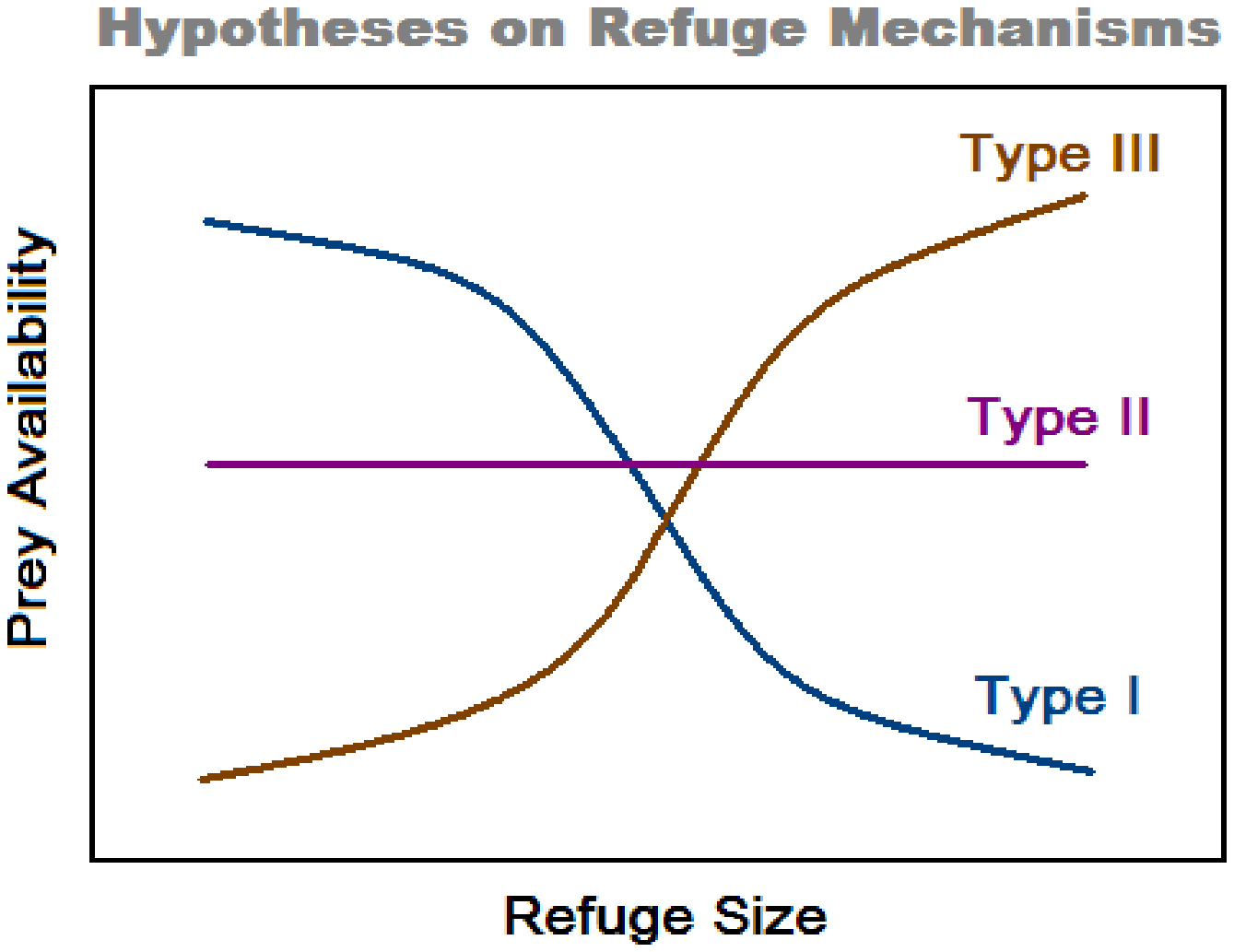,height=3in,width=4in,angle=0}}}
\caption{\label{Hypotheses} Three possible biological hypotheses for
the effects of the refuge size on the prey availability for
predators.}
\end{figure}

Figure \ref{Hypotheses}. Three biological hypotheses for
the effects of the refuge size on the prey availability for
predators. Type I: the prey availability for predators is a
decreasing function of the refuge size, because the refuge provides
places for prey to hide from predators. Type II: the prey
availability for predators is independent of the refuge size in the
sense of density (per unit area), because in a number of cases prey
biomass is proportional to the refuge size. Type III: the prey
availability for predators is an increasing function of the refuge
size, because the refuge both provides prey to predators and stores
prey for latter consumption by predators.

\newpage

\begin{thebibliography}{}

\bibitem[Berger(1991)]{ber91}
Berger, J., 1991. Pregnancy incentives, predation constraints and
habitat shifts: experimental and field evidence for wild bighorn
sheep. Anim. Behav. 41, 61-77.

\bibitem[Cassini(1991)]{cas91}
Cassini, M.H., 1991. Foraging under predation risk in the wild
guinea pig \textit{Cavia aperea}. Oikos 62, 20-24.

\bibitem[Clarke et al.(1993)]{cla93}
Clarke, M.F., da Silva, K.B., Lair, H., Pocklington, R., Kramer,
D.L., and Mclaughlin, R.L., 1993. Site familiarity affects escape
behaviour of the eastern chipmunk, \textit{Tamius striatus}. Oikos
66, 533-537.

\bibitem[Cowlishaw(1997)]{cow97}
Cowlishaw, G., 1997. Refuge use and predation risk in a desert
baboon population. Anim. Behav. 54, 241-253.

\bibitem[Dash(2001)]{das01}
Dash, M.C., 2001. Fundamentals of Ecology. Tata McGraw-Hill.

\bibitem[Del Giorgia et al.(1999)]{del99}
Del Giorgia, P.A., Cole, J.J., Caraco, N.F., and Peters, R.H., 1999.
Linking Planktonic Biomass and Metabolism to Net Gas Fluxes in
Northern Temperate Lakes. Ecology 80, 1422-1431.

\bibitem[Dill and Houtman(1989)]{dil89}
Dill, L.M. and Houtman, R., 1989. The influence of distance to
refuge on flight-initiation distance in the grey squirrel
(\textit{Sciurus carolinensis}). Can. J. Zool. 67, 232-235.

\bibitem[Doherty and Sale(1985)]{doh85}
Doherty, P.J. and Sale, P.F., 1985. Predation on juvenile coral reef
fishes: and exclusion experiment. Coral Reefs 4, 225-234.

\bibitem[Friedlander and Martini(2002)]{fri02}
A.M. Friedlander and Martini E.E., 2002. Contrasts in density, size, and
biomass of reef fishes between the northewestern and the main Hawaiian
islands: the effects of fishing down apex predators. Marine Ecology
Progress Series 230, 253-264.

\bibitem[Gonzalez-Olivares and Ramos-Jiliberto(2003)]{gon03}
Gonzalez-Olivares, E. and Ramos-Jiliberto, R., 2003. Dynamic consequences of prey refuges in a simple system: more prey, fewer predators and enhanced stability. Ecological modeling 166, 135-146.

\bibitem[Hausrath(1994)]{hau94}
Hausrath, A., 1994. Analysis of a model predator-prey system with
refuges. J. Math. Anal. Appl. 181, 531-545.

\bibitem[Hawkins et al.(1993)]{haw93}
Hawkins, B.A., Thomas, M.B., and Hochberg, M.E., 1993. Refuge theory
and biological control. Science 262, 1429-1432.

\bibitem[M. Hay, pers. comm.(2008)]{hay08}
M. Hay, 2008. Personal Communication.

\bibitem[Hixon and Beets(1993)]{hix93}
Hixon, M.A. and Beets, J.P., 1993. Predation, Prey Refuges, and the
Structure of Coral-Reef Fish Assemblages. Ecological Monographs 63,
77-101.

\bibitem[Holling(1959a)]{hol59a}
Holling, C.S., 1959a. The components of predation as revealed by a
study of small mammal predation of the European Pine Sawfly.
Canadian Entomologist 91, 293-320.

\bibitem[Holling(1959b)]{hol59b}
Holling, C.S., 1959b. Some characteristics of simple types of
predation and parasitism. Canadian Entomologist 91, 385-398.

\bibitem[Holmes(1991)]{hol91}
Holmes, W.G., 1991. Predator risk affects foraging piks:
observational and experimental evidence. Anim. Behav. 42, 111-119.

\bibitem[Huang et al.(2006)]{hua06}
Huang, Y., Chen, F., and Zhong, L., 2006. Stability analysis of a
prey-predator model with holling type III response function
incorporating a prey refuge. Applied Mathematics and Computation
182, 672-683.

\bibitem[Huffaker(1958)]{huf58}
Huffaker, C.B., 1958. Experimental studies on predation: dispersion
factors and predator-prey oscillations. Hilgardia 27, 343-383.

\bibitem[Kar(2005)]{kar05}
Kar, T.K., 2005. Stability analysis of a prey-predator model
incorporation a prey refuge. Commun. Nonlinear Sci. Numer. Simul.
10, 681-691.

\bibitem[Kar(2006)]{kar06}
Kar, T.K., 2006. Modelling and analysis of a harvested prey-predator
system incorporating a prey refuge. Journal of Computational and
Applied Mathematics 185, 19-33.

\bibitem[Ko and Ryu(2006)]{ko06}
Ko, W. and Ryu, K., 2006. Qualitative analysis of a predator-prey
model with Holling type II functional response incorporating a prey
refuge. J. Differential Equations 231, 534-550.

\bibitem[Legrand and Barbosa(2003)]{leg03}
Legrand, A. and Barbosa, P., 2003. Plant morphological complexity
impacts foraging efficiency of adult Coccinella septempunctata L.
(Coleoptera: Coccinellidae). Environ. Entomol. 32, 1219-1226.

\bibitem[Lima and Dill(1990)]{lim90}
Lima, S.L. and Dill, L.M., 1990. Behavioral decisions made under the risk of predation: a review and prospectus.  Canadian Journal of Zoology 68, 619-640.

\bibitem[Lotka(1925)]{lot25}
Lotka, A.J., 1925. Elements of Physical Biology. Williams and
Wilkins, Baltimore.

\bibitem[Lubow and Smith(2004)]{lub04}
Lubow, B.C. and Smith, B.L., 2004. Population dynamics of the Jackson elk herd. Journal of Wildlife management 68, 810-829.

\bibitem[MacHutchon and Harestad(1990)]{mac90}
MacHutchon, A.G. and Harestad, A.S., 1990. Vigilance behaviour and
use of rocks by Columbian ground squirrels. Canadian Journal of
Zoology 68, 1428-1433.

\bibitem[McNair(1986)]{mcn86}
McNair, J., 1986. The effects of refuges on predator-prey
interactions: A reconsideration. Theoret. Population Biol. 29,
38-63.

\bibitem[Murdoch and Oaten(1975)]{mur75}
Murdoch, W.W. and Oaten, A., 1975. Predation and Popluation
Stability. Advances in Ecological Research 9, 1-131.

\bibitem[Odum(1971)]{odu71}
Odum, E.P., 1971. Fundamentals of Ecology. W.B Saunders,
Philadelphia, Pennsylvania, USA.

\bibitem[Pauly and Christensen(1995)]{pau95}
Pauly, D. and Christensen, V., 1995. Primary production required to
sustain global fisheries. Nature 374, 255-257.

\bibitem[Persson and Eklov(1995)]{per95}
Persson, L. and Eklov, P., 1995. Prey refuges affecting interactions between piscivorous perch and juvenile perch and roach. Ecology 76, 70-81.

\bibitem[Reaney(2007)]{rea07}
Reaney, L.T., 2007. Foraging and mating opportunities influence refuge use in the fiddler crab, Uca mjoebergi. Animal Behaviour 73, 711-716.

\bibitem[Reichle(1981)]{rei81}
Reichle, D.E., 1981. Dynamic Properties of Ecosystems. Cambridge
University Press, New York, USA.

\bibitem[Rossi et al.(2006)]{ros06}
Rossi, M.N., Reigada, C., and Godoy, W.A.C., 2006. The role of
habitat heterogeneity for the functional response of the spider
Nesticodes rufipes (Araneae: Theridiidae) to houseflies. Appl.
Entomol. Zool. 41, 419-427.

\bibitem[Sandin et al.(2008)]{san08}
Sandin, S.A., Smith, J.E., DeMartini, E.E., Dinsdale, E.A., Donner,
S.D., Fiedlander, A.M., et al., 2008. Baselines and Degradation of
Coral Reefs in the Northern Line Islands. PLoS ONE 3, e1548.

\bibitem[Shulman(1984)]{shu84}
Shulman, M.J., 1984. Resource limitation and recruitment patterns in
a coral reef fish assemblage. J. Exp. Mar. Biol. Ecol. 74, 85-109.

\bibitem[Sih(1987)]{sih87}
Sih, A., 1987. Prey refuges and predator-prey stability. Theoret.
Population Biol. 31, 1-12.

\bibitem[Sih(1997)]{sih97}
Sih, A., 1997. To hide or not to hide? Refuge use in a fluctuating
environment. Trends in Ecology \& Evolution 12, 375-376.

\bibitem[Singh et al.(2008)]{sin08}
Singh, A., Wang, H., Morrison, W., and Weiss, H., 2008. Fish Biomass
Structure at Pristine Coral Reefs and Degradation by Fishing.
Manuscript.

\bibitem[Smith, pers. comm.(2008)]{smi08}
Smith, B., 2008. Personal Communication.

\bibitem[Volterra(1926)]{vol26}
Volterra, V., 1926. Variazioni e fluttuazioni del numero d'individui
in specie animali conviventi. Mem. R. Accad. Naz. dei Lincei. Ser.
VI, Vol. 2.

\bibitem[Warburton et al.(1998)]{war98}
Warburton, K., Retif, S., and Hume, D., 1998. Generalists as sequential specialists: diets and prey switching in juvenile silver perch. Environmental Biology of fishes 51, 445-454.

\bibitem[Ward et al.(2001)]{war01}
Ward T.J., Heinemann, D., and Evans, N., 2001. The Role of Marine Reserves as Fisheries Management Tools: a review of concepts, evidence and international experience. Bureau of Rural Sciences, Canberra, Australia.

\bibitem[White and Andow(2007)]{whi07}
White, J.A. and Andow, D.A., 2007. Foraging for intermittently
refuged prey: theory and field observations of a parasitoid. Journal
of Animal Ecology 76, 1244-1254.

\bibitem[Inverse function in Wikipedia(internet)]{wik}
Wikipedia. Link: http://en.wikipedia.org/wiki/Inverse\_function.

\end{thebibliography}
\end{document}